\documentclass{emulateapj}
\usepackage{amsmath}
\def\H{H_0 = 70~\rm km~s^{-1}~Mpc^{-1}}

\begin{document}
\title{
The Spatial Structure of An Accretion Disk
}

\author{Shawn Poindexter\altaffilmark{1},
	Nicholas Morgan\altaffilmark{1},
	Christopher S. Kochanek\altaffilmark{1}
}
\altaffiltext{1}{Department of Astronomy, Ohio State University, 
140 West 18th Avenue, Columbus, OH 43210, USA,
(sdp,nmorgan,ckochanek)@astronomy.ohio-state.edu}

\begin{abstract}

Based on the microlensing variability of the two-image gravitational lens
HE~1104--1805 observed between $0.4$ and $8~\mu$m, we have measured the
size and wavelength-dependent structure of the quasar accretion disk.
Modeled as a power law in temperature, $T\propto R^{-\beta}$,
we measure a 
B-band ($0.13~\mu$m in the rest frame) half-light radius of
$R_{1/2,\rm B} = 6.7^{+6.2}_{-3.2}\times10^{15}~$ cm ($68\%$ CL)
and a logarithmic slope of
$\beta=0.61^{+0.21}_{-0.17}$ ($68\%$ CL)
for our standard model with a logarithmic prior on the disk size.
Both the scale and the slope are consistent with simple thin disk models where
$\beta=3/4$ and $R_{1/2,\rm B} = 5.9\times 10^{15}~\rm cm$ for a
Shakura-Sunyaev disk radiating at the Eddington limit with $10\%$ efficiency.
The observed fluxes favor a slightly shallower slope,
$\beta=0.55^{+0.03}_{-0.02}$, and a significantly smaller size for
$\beta=3/4$.

\end{abstract}


\section{Introduction}\label{sec:intro}

A simple theoretical prediction for thermally radiating
thin accretion disks well outside the inner disk edge
is that the temperature diminishes with radius as
$T \propto R^{-3/4}$ \citep{Shakura73}.
This implies a characteristic size at wavelength $\lambda$ of
$R_\lambda \propto \lambda^{4/3}$, where $R_\lambda$ is the radius
at which $kT = hc/\lambda$.
Needless to say, it is unlikely that disks are this simple \citep[e.g.][]{Blaes04},
but measurement of the size-wavelength scaling would be a fundamental
test for any disk theory.
While the angular sizes of quasar accretion disks are far too small
to be resolved by direct observation, gravitational lenses can
serve as natural telescopes to probe accretion disk structure on
these scales.

Here we make the first such measurement of this size-wavelength scaling.
Our approach, gravitational microlensing of a quasar, will be unfamiliar to the
AGN community, but it is the only method with the necessary spatial
resolution available to us for the foreseeable future
\citep[see the review by][]{Wambsganss06}.
The stars in the lens galaxy near the image of a multiply-imaged quasar generate
complex caustic networks with a characteristic scale
called the Einstein radius $R_{\rm E}$ set by the
mean stellar mass $\left<M\right>$.
For the lens system we consider here, HE~1104--1805,
$\left<R_{\rm E}\right> = 3.6\times10^{16}(\left<M\right>/h M_\sun)^{1/2}~\rm cm$.
Because the magnification diverges on the caustic curves of the pattern
and the source is moving relative to the lens and observer,
microlensing allows us to study the spatial structure of
anything smaller than $\left<R_{\rm E}\right>$.
Using microlensing to probe scales smaller than $R_{\rm E}$
has been successfully applied to resolve stars in Galactic
microlensing events \citep[e.g.][]{Albrow01}.
For quasars, this has been considered analytically or with simulations
\citep*[e.g.][]{Agol99,Goicoechea04,Grieger91},
but the data and algorithms needed to implement
the programs have only become available recently \citep[see][]{Kochanek04}.

HE~1104--1805 is a doubly imaged radio-quiet quasar 
at $z_s=2.319$ with a separation of $3\farcs15$ \citep{Wisotzki93}.
The lens at $z_l = 0.729$ was discovered in the near-IR by \citet*{Courbin98}
and with {\it HST} \citep{Remy98,Lehar00}.
Here we analyze 13 years of photometric data in 11 bands from the
mid-IR to B-band using the methods of \citet{Kochanek04} to measure the
wavelength-dependent structure of this quasar modeled as a power law
$R_\lambda \propto \lambda^{1/\beta}$.
We assume a flat $\Lambda$CDM cosmological model with
$\Omega_M = 0.3$ and $\H$ and
report the disk sizes assuming a mean inclination of $\cos{i}=1/2$.
In \S\ref{sec:datamethod} we describe the data set and our methods, 
and \S\ref{sec:results} presents our measurement results and
our conclusions.

\section{Data and Methods}\label{sec:datamethod}

\begin{figure}[t]
\plotone{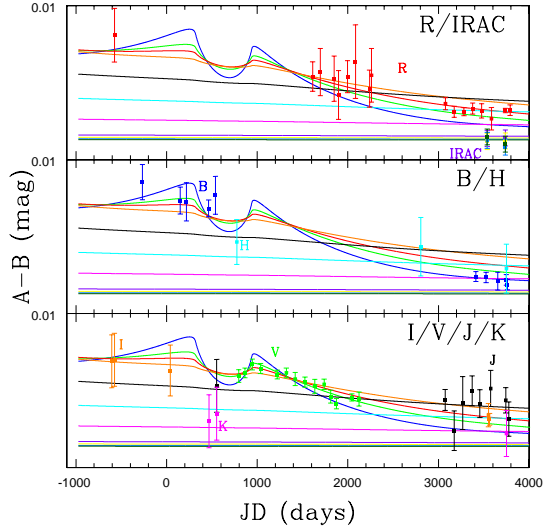}
\caption{The HE~1104--1805 multi-band light curves.
The curves show a model for the BVRIJHK bands and the four IRAC channels
($3.6$ to $8.0~\mu$m).
For clarity we split the data into three panels and
show the B, V, R, and J-band data points in 200 day averages.
\label{fig:lc}}
\end{figure}

We included observations of HE~1104--1805 in 11 bands:
B, V, R, I, J, H, K, $3.6~\mu{\rm m}$, $4.5~\mu{\rm m}$,
$5.8~\mu{\rm m}$, and $8.0~\mu{\rm m}$.
These included our own SMARTS optical/near-IR, SOAR near-IR,
{\it HST} and Spitzer IRAC data \citep{Poindexter07},
R-band monitoring data by \citet{Ofek03},
the V-band monitoring data from \citet{Schechter03} and \citet{Wyrzykowski03},
and earlier data from \citet{Remy98}, \citet*{Gil-Merino02},
\citet{Courbin98}, and \citet{Lehar00}.
Where possible we corrected the light curves for the 152 day time delay
between the images we measured in \citet{Poindexter07}.
Where we could not, we broadened the photometric
uncertainties by 0.07 mag so that the flux ratio uncertainties
would be larger by the 0.1 mag shifts we found between time-delay corrected
and uncorrected flux ratios.
The light curve is plotted along with one of the light curve models from our
analysis in Figure \ref{fig:lc}.  As pointed out in \citet{Poindexter07}, 
image A has slowly switched from being bluer than image B 
to being redder in the optical/near-infrared (see Figure \ref{fig:lc}),
while the mid-infrared flux agrees with the flux ratio of the 
broad emission lines \citep{Wisotzki93}. 

Determining the structure of the disk as a function of wavelength from
such data is relatively straight forward.  The divergences
on the caustics of the microlensing magnification patterns are only
renormalized by the finite size of the source quasar because 
the observed magnification is the convolution of the pattern with the 
source structure.  Thus, larger emission regions will show smaller 
variability amplitudes than smaller emission regions because they
smooth the patterns more.  As we go from the K-band to the
B-band, the radius of a standard thin disk with $R_\lambda \propto \lambda^{4/3}$,
changes by a factor of $8.6$, corresponding to a change in the
disk area of almost two orders of magnitude.  We see in Fig.~\ref{fig:lc}
that the bluer wavelengths show larger amplitudes than the red wavelengths,
so we immediately know that the blue emission regions are more compact
than the red.

Our analysis uses the Bayesian Monte Carlo method of \citet{Kochanek04}
to analyze the data.
In essence, we randomly draw large numbers of trial light curves 
from a range of physical models, fit them to the data and then use Bayesian 
statistics to derive probability distributions for the disk structure. 
We need, however, to discuss the physical variables used in the models
over which we average as well as our model for the structure of the 
accretion disk.

We use the lens model sequence from \citet{Poindexter07}, which consists
of a de Vaucouleurs model matched to the {\it HST} observations
embedded in an NFW halo, to set the shear $\gamma$, convergence $\kappa$, and
stellar fraction $\kappa_*/\kappa$ for the microlensing magnification patterns.
The models were constrained to match the flux ratios of the mid-IR IRAC bands.
We used a mass function of $dN/dM \propto M^{-1.3}$ with $M_{max}/M_{min} = 50$
whose structure is broadly consistent with the Galactic disk mass function
of \cite{Gould00}. 
The lens models were parameterized by $f_{M/L}$, the fraction of a
constant mass-to-light ratio ($M/L$) represented by the visible galaxy.
For each of ten models, $f_{M/L} = 0.1,0.2,\ldots,1.0$, we produce 2
magnification patterns with an outer dimension of $10 R_{\rm E}$ and
$8192^2$ pixels to achieve a pixel scale of
$4.4\times 10^{13}~(\left<M\right>/hM_\sun)^{1/2}$ cm/pixel
that is smaller than the gravitational radius
$r_g = GM/c^2 = 3.5\times 10^{14}~\rm cm$ expected for HE~1104--1805
(see \S\ref{sec:results}).
We experimented extensively with magnification patterns of different sizes
and scales to ensure that our choice of pixel scale and outer dimension did not
affect our results.
We used the velocity model from \citet{Kochanek04}, where the adopted values
are $73~\rm km/s$ for our velocity projected onto the lens plane,
$308~\rm km/s$ for the velocity dispersion of the lens estimated from
fitting an isothermal lens model and RMS peculiar velocities of
$135$ and $71~\rm km/s$ for the lens and source respectively.
We assume a uniform prior on the mean masses of the stars of
$0.1~M_\sun < \left<M\right> < 1.0~M_\sun$, but this has only
limited effects on the estimates of the disk size (see Kochanek 2004).
Our estimates of the disk properties average over all these parameters.

\begin{figure}
\plotone{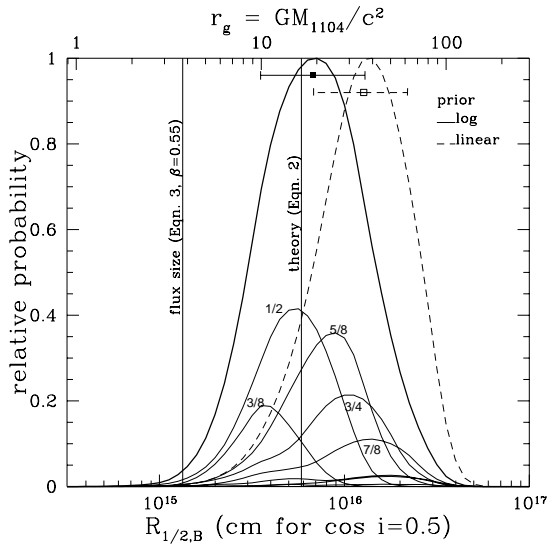}
\caption{Probability distributions for the B-band accretion disk half-light radius.
The bold solid (dashed) curve is the distribution based on a logarithmic
(linear) prior on the disk size.
The filled (open) square near the top is the median of the distribution
with a logarithmic (linear) prior along with the $68\%$ confidence error bars.
The lower curves show the contribution from the more significant
$\beta$ trials, labeled by their value of $\beta$.
The vertical lines show the
B-band thin disk size predicted from the I-band flux (Eqn. \ref{eqn:fluxsize})
and the size expected from standard thin disk theory (Eqn.
\ref{eqn:theorysize}) for Eddington limited ($L/L_{\rm E}=1$)
accretion with $\eta=10\%$ efficiency.
\label{fig:BbandSize}}
\end{figure}

We modeled the disk as a face-on thermally radiating disk with temperature
$T\propto R^{-\beta}$ \citep[e.g.][]{Collier98},
corresponding to a surface brightness profile of
\begin{equation}
f_\nu \propto \nu^3 \left[\exp{(R/R_\lambda)}^\beta-1\right]^{-1}
\label{eqn:fnu}
\end{equation}
where the size scale is
$R_\lambda=R_{\rm B}(\lambda/\lambda_{\rm B})^{1/\beta}$
and $R_{\rm B}$ is the disk size at the observed
B-band ($1310$\AA~in the rest frame) and $\beta=3/4$ for standard
thin disk theory.
In many cases we report the half-light radius
$R_{1/2,\lambda}(\beta) = C(\beta)R_\lambda(\beta)$ where
for $\beta=3/4$, $C=2.44$.
While we do not include an inner disk edge of $R_{in}\simeq 2r_g$ 
to $6r_g$, the lack of this central hole has little effect
on our results unless $\beta\rightarrow 2$ or $R_B\simeq R_{in}$.
We tried disk models with profile exponents of $\beta=1/4,3/8,\cdots,7/4$,
making $4 \times 10^{6}$ trial light curves for each value of $\beta$.
In our final analysis we use the $3.3\times10^6$ light curves that
passed a threshold of $\chi^2/N_{dof} \leq 3$ for $N_{dof}=438$
degrees of freedom.  We used both a logarithmic prior,
$P(R_\lambda) \propto 1/R_\lambda$, and a linear prior,
$P(R_\lambda) \propto \rm constant$, on the disk size.
Generally, a logarithmic prior is preferred for scale free
variables like $R_\lambda$,
but for this problem a linear prior may be more appropriate for small
source sizes because we are sensitive to the difference between small sizes only
during caustic crossings.  If, however, we have solutions with the caustic
crossing sitting in a gap of the light curve, we will find that
$P(D|R_\lambda)$ approaches a constant value as $R_\lambda$ goes to zero.
This leads to a formal divergence in $P(R_\lambda|D)$ with a logarithmic
prior, suggesting that a linear prior may be more appropriate.  In practice,
Figs. \ref{fig:BbandSize}-\ref{fig:BbandVsBeta} show the results for both
options and we use the results
for the logarithmic prior in our discussion. 
For the results we give the value at the median of the probability 
distribution and the 68\% ($1~\sigma$) confidence regions.

\section{Results and Discussion}\label{sec:results}

We have no difficulty reproducing the observed light curves, including
the presently observed color reversal.  We illustrate this in 
Fig.~\ref{fig:lc}, where we superpose our best fitting light
curve model on the data.

\begin{figure}
\plotone{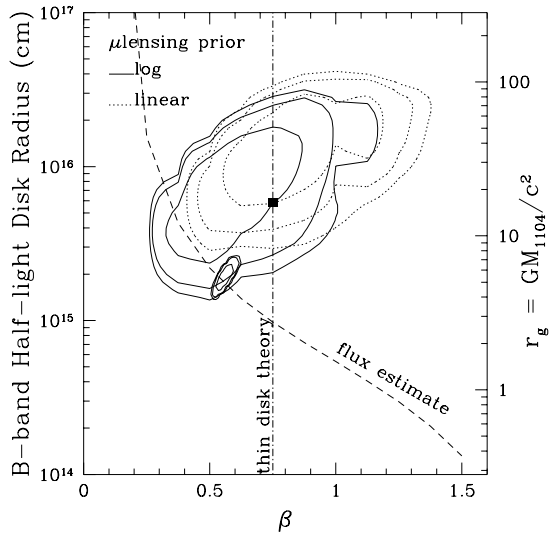}
\caption{B-band half-light disk radius $R_{1/2,\rm B}$
versus $\beta$ where $T\propto R^{-\beta}$.
The solid (dotted) contours represent the $68\%$, $90\%$, and $95\%$ confidence
levels of the microlensing measurements assuming a logarithmic (linear) prior on disk size.
The small solid contours is the B-band flux estimate considering a
$T\propto R^{-\beta}$ model fit to all the bands.
The dashed line shows the size estimated from just the B-band flux
measurement without the constraint on $\beta$ from the other bands.
The large solid square at $\beta=3/4$ shows the radius
predicted by thin disk theory (Eqn. \ref{eqn:theorysize}).
\label{fig:BbandVsBeta}}
\end{figure}

\begin{figure}
\plotone{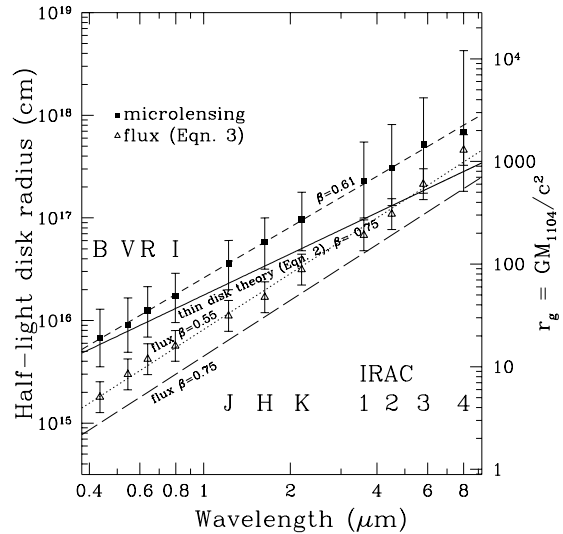}
\caption{Half-light disk radius $R_{1/2,\lambda}$ versus wavelength.
The filled squares (open triangles) are the size estimates from microlensing
(the source flux, Eqn. \ref{eqn:fluxsize}).
Note that the microlensing uncertainties are highly correlated
because the only 2 actual variables are $R_{\rm B}$ and $\beta$.
The short dashed and dotted lines are the best fit power laws to these measurements.
The long dashed line shoes how the normalization of the points would shift
if we use $\beta=3/4$ rather than the best fit slope of $\beta=0.55$.
\label{fig:sizeVsWavelength}}
\end{figure}

We find a wavelength-size scaling of the accretion disk of
$\beta=0.61^{+0.21}_{-0.17}$ and $\beta=0.89^{+0.23}_{-0.16}$
for the logarithmic and linear priors respectively 
(Fig.~\ref{fig:sizeVsWavelength}), where $R_\lambda \propto\lambda^{1/\beta}$
and $T\propto R^{-\beta}$.  This is consistent with simple
thin disk theory ($\beta=3/4$).  The B-band half light 
radius is $R_{1/2,\rm B} = 6.7^{+6.2}_{-3.2}\times10^{15}$~cm 
with the logarithmic prior and
$R_{1/2,\rm B} = 1.3^{+0.9}_{-0.6}\times10^{16}$~cm
with the linear prior (Figs. \ref{fig:BbandSize} and \ref{fig:BbandVsBeta}).
We use the half light radius rather than $R_\lambda$ because it has
less covariance with the exponent $\beta$.
We can also estimate the sizes for the individual bands, as shown in
Fig.~\ref{fig:sizeVsWavelength}, although these will be highly 
correlated because the model only has the exponent $\beta$
and one scale length as actual parameters.  If we fix
$\beta=3/4$, then we find that B-band size for this case is
$R_{1/2,\rm B}(\beta=3/4) = 9.0^{+6.5}_{-4.2}\times10^{15}$ cm.

We can compare to thin disk theory only for the case of $\beta=3/4$ since
there is no simple generalization for alternate values of $\beta$.  Based 
on the CIV emission line width of HE~1104--1805, \citet{Peng06} estimated 
a black hole (BH) mass of $M_{1104} = 2.4\times10^9 M_\sun$, which corresponds 
to a gravitational radius of $r_g = GM/c^2 = 3.5\times 10^{14}~\rm cm$.
In thin disk theory \citep{Shakura73}, this implies a size scaling of
\begin{equation}
\begin{split}
R_\lambda &= \frac{1}{\pi^2}\left[\frac{45}{16}\frac{\lambda_{rest}^4 r_g \dot M}{h_p}\right]^{1/3} \\
&= (1.7\times10^{16}) \left[\frac{\lambda_{rest}}{\mu\rm m}\right]^{4/3}
\left[\frac{M_{BH}}{M_{1104}}\right]^{2/3}
\left[\frac{L}{\eta L_{\rm E}}\right]^{1/3}~{\rm cm},
\end{split}
\label{eqn:theorysize}
\end{equation}
where $\eta$ is the radiative efficiency ($L=\eta\dot M c^2$), $h_p$ is
the Planck constant,
and $L/L_{\rm E}$ is the fraction
of the Eddington luminosity radiated by the QSO.  Thus the expected B-band
(rest frame $0.13~\mu$m) size for a disk radiating at the Eddington limit 
($L/L_{\rm E}$) with 10\% efficiency is $R_{\rm B} = 2.4\times10^{15}$ cm 
corresponding to a half-light radius of $R_{1/2,\rm B} = 5.9\times10^{15}$ cm. 
This agrees well with our measurement (Figs.~\ref{fig:BbandSize},
\ref{fig:BbandVsBeta}, and \ref{fig:sizeVsWavelength}).
For the $\beta=3/4$ model the disk scale
length is significantly larger than the gravitational radius  
($R_{1/2,\rm B} \sim 20 r_g$) and corrections for the inner edge of the
disk will be modest.  Particularly if we allow for uncertainties
in the black hole mass estimate ($\sim 0.3$~dex), there is good agreement 
with the simplest possible thin disk model.   

The magnification-corrected flux of the quasar provides a second
comparison scale under the assumption that the disk is thermally
radiating.  If the quasar has a magnification corrected magnitude of
$m$, then 
\begin{equation}
\begin{split}
R_\lambda(\beta) \simeq \frac{2.8\times10^{15}}{h \sqrt{K(\beta) \cos i}}
                \frac{D_{OS}}{r_H}
		\left[\frac{\lambda_{obs}}{\mu\rm m}\right]^{3/2} \\
                \left[{\frac{\hbox{zpt}}{2409~\hbox{Jy}}}\right]^{1/2}
		10^{-0.2({\rm m}-19)}  \hbox{cm}.
\label{eqn:fluxsize}
\end{split}
\end{equation}
where $\lambda_{obs}$ is the observed wavelength, $\hbox{zpt}$
is the filter zero point (normalized to the I-band), and
\begin{equation}
   K(\beta) = \frac{1}{2.58}\int_0^\infty u du \left[ \exp(u^\beta)-1\right]^{-1}
\end{equation}  
is the $\beta$-dependent term due to the temperature profile normalized
so that $K(\beta=3/4)=1$.
Assuming the magnifications of images A and B are
11.5 and 4 respectively (as found
in the best fit macro model in \citealt{Poindexter07}),
we calculated the magnification corrected source magnitude in
each of the 11 bands to estimate
the disk size versus wavelength (Fig. \ref{fig:sizeVsWavelength}).
For the H-, I-, and V-band, we used
the {\it HST} observations of \citet{Lehar00}.
The mid-IR magnitudes are from {\it Spitzer} \citep{Poindexter07}.
We calibrated the SMARTS B/R and J/K data using the Guide Star
Catalog and 2MASS respectively.

The sizes estimated from the flux are very well fit by
a power law (see Fig. \ref{fig:sizeVsWavelength})
with a slope equivalent
to $\beta=0.55^{+0.03}_{-0.02}$ when we assume a magnification uncertainty of
a factor of 2.  While this slope is consistent with our
microlensing results, the size scale of
$R_{1/2,\rm B} = 1.8^{+0.7}_{-0.5}\times10^{15}$ is smaller
by a factor of 4 than
our standard estimate.  This discrepancy depends on the
value of $\beta$, with a flatter temperature profile showing
less of a difference.
While there is little extinction in the lens, 2.9 mags of B-band
($0.31~\mu$m rest frame) extinction in the source would reconcile
the flux and microlensing size estimates.  However, with such an
extinction, neither SMC \citep{Gordon03} or AGN
\citep{Gaskell04} extinction curves can reconcile the two
estimates at all wavelengths.  Nonetheless absorption in the source
could be a partial explanation.

The general relationship we find between the microlensing,
thin disk and flux sizes seems to be typical.  \citet{Pooley07}
noted qualitatively that microlensing sizes tended to be larger
than expected from the optical flux and thin disk models.
\citet{Morgan07} showed quantitatively that the microlensing sizes scaled as
expected with BH mass ($R_\lambda \propto M^{2/3}$) and were
consistent with being proportional to the flux sizes, but that the
absolute scales of the microlensing sizes were slightly larger
than the thin disk sizes and considerably larger than the
flux sizes.
Our results here suggest that part of the solution
may be that the effective temperature profile is somewhat
shallower than $T \propto R^{-3/4}$.

Several local estimates \citep{Collier99,Sergeev05} have found
UV-optical wavelength dependent time delays of nearby AGN consistent
with $T\propto R^{-3/4}$.  They also found the flux discrepancy, but
phrased the problem as needing to put the systems at higher than
expected distances (through a low value of $H_0$) in order to reconcile
the model disk surface brightness with the observed flux.

Our current disk model is a face-on, thermally radiating disk without
the central temperature depression created by the inner edge of the
accretion disk (Eqn \ref{eqn:fnu}).
Omitting the inner edge has little effect because
at fixed wavelength little flux is radiated there and the finite
resolution of the magnification patterns
eliminates the formal surface brightness divergence in Eqn. (\ref{eqn:fnu}).
It is expected from theoretical work that microlensing
is primarily sensitive to an effective smoothing area, and the resulting
size estimate is only weekly sensitive to the true surface brightness
profile \citep[e.g.][]{Mortonson05, Congdon07}.
Nonetheless, a clear next step is to begin interpreting the 
microlensing results using more realistic thin disk
models \citep[e.g.][]{Hubeny01} to see if adding such details
alters the basic picture.  The present results suggest that part
of the solution may be to alter the temperature profile of the disk,
perhaps through irradiation of the outer regions by the inner regions.

\acknowledgments
We thank R. Pogge, E. Agol, C. Morgan, and X. Dai for discussions
on microlensing, accretion disks, and suggestions for improving the
manuscript.
Support for this work was provided by NASA through an award
GRT00003172 issued by JPL/Caltech.
This work is based in part on observations made with the
Spitzer Space Telescope,
which is operated by the Jet Propulsion Laboratory, California Institute of
Technology under a contract with NASA.
Based on observations made with the NASA/ESA Hubble Space Telescope, obtained
at the Space Telescope Institute. STScI is operated by
the association of Universities for Research in Astronomy, Inc. under the NASA
contract NAS5-26555.
This publication makes use of data products from the Two Micron All Sky Survey,
which is a joint project of the University of Massachusetts and the Infrared
Processing and Analysis Center/California Institute of Technology, funded by
NASA and NSF.
The Guide Star Catalogue-II is a joint project of the STScI 
and the Osservatorio Astronomico di Torino. STScI
is operated by the AURA, for NASA 
under contract NAS5-26555. The participation
of the Osservatorio Astronomico di Torino is supported by the Italian
Council for Research in Astronomy. Additional support is provided by
ESO, Space Telescope European Coordinating
Facility, the International GEMINI project and the ESA
Astrophysics Division.

\end{document}